\documentclass[aps,showpacs,amsmath,amssymb,12pt]{revtex4}
\usepackage{latexsym}
\usepackage{bm}

\begin{document}

\title{G{\" o}del-type Metrics in Various Dimensions II: Inclusion 
of a Dilaton Field}
\author{Metin G{\" u}rses}
\email{gurses@fen.bilkent.edu.tr}
\affiliation{Department of Mathematics, Faculty of Sciences,\\
             Bilkent University, 06800, Ankara, Turkey}

\author{{\" O}zg{\" u}r Sar{\i}o\u{g}lu}  
\email{sarioglu@metu.edu.tr}
\affiliation{Department of Physics, Faculty of Arts and  Sciences,\\
             Middle East Technical University, 06531, Ankara, Turkey}

\date{\today}

\begin{abstract}
This is the continuation of an earlier work where G\"{o}del-type metrics
were defined and used for producing new solutions in various dimensions.
Here a simplifying technical assumption is relaxed which, among other
things, basically amounts to introducing a dilaton field to the models
considered. It is explicitly shown that the conformally transformed 
G\"{o}del-type metrics can be used in solving a rather general class 
of Einstein-Maxwell-dilaton-3-form field theories in $D \geq 6$ dimensions. 
All field equations can be reduced to a simple ``Maxwell equation'' in 
the relevant $(D-1)$-dimensional Riemannian background due to a neat 
construction that relates the matter fields. These tools are then used in 
obtaining exact solutions to the bosonic parts of various supergravity 
theories. It is shown that there is a wide range of suitable backgrounds 
that can be used in producing solutions. For the specific case of 
$(D-1)$-dimensional trivially flat Riemannian backgrounds, the 
$D$-dimensional generalizations of the well known Majumdar-Papapetrou 
metrics of general relativity arise naturally.
\end{abstract}


\maketitle

\section{\label{intro} Introduction}

Let $M$ be a $D$-dimensional manifold with a metric of the form
\begin{equation}
g_{\mu \nu}= h_{\mu \nu}-u_{\mu} u_{\nu} \; . \label{met}
\end{equation}
(We take Greek indices to run from 0,1, $\cdots$ to $D-1$ and our conventions 
are similar to the conventions of Hawking-Ellis \cite{eh}.)
Here $h_{\mu \nu}$ is a degenerate $D \times D$ matrix with rank equal to
$D-1$. We assume that the degeneracy of $h_{\mu\nu}$ is caused by 
taking $h_{k\mu}=0$, where $x^{k}$ is a fixed coordinate with 
$0 \le k \le D-1$ (note that $x^{k}$ does not necessarily have to 
be spatial), and that the rest of $h_{\mu\nu}$, i.e. $\mu \neq k$ 
or $\nu \neq k$, is dependent on all the coordinates $x^{\alpha}$ 
except $x^{k}$ so that $\partial_{k} h_{\mu\nu}=0$. Hence, in the most general
case, ``the background'' $h_{\mu\nu}$ can effectively be thought of as the 
metric of a $(D-1)$-dimensional non-flat spacetime. As for $u^{\mu}$,
we assume that it is a timelike unit vector, $u_{\mu} u^{\mu}=-1$, and 
that $u_{\mu}$ is independent of the fixed special coordinate $x^{k}$, i.e. 
$\partial_{k} u_{\mu}=0$. The assumptions so far imply that the determinant 
of $g_{\mu\nu}$ is $g=-u_{k}^{2}\, h$, where $h$ is the determinant of 
the $(D-1)\times(D-1)$ submatrix obtained by deleting the $k^{th}$ row 
and the $k^{th}$ column of $h_{\mu\nu}$, and moreover 
$u^{\mu}=-\frac{1}{u_{k}}\, \delta^{\mu}_{\;k}$. 

The question we ask now is the following: Let us start with a metric of the
form (\ref{met}) and calculate its Einstein tensor. Can the Einstein
tensor be interpreted as describing the energy momentum tensor of a 
physically acceptable source? Does one need further assumptions on
$h_{\mu\nu}$ and/or $u_{\mu}$ so that ``the left hand side'' of 
$G_{\mu\nu} \sim T_{\mu\nu}$ can be thought of giving an acceptable
``right hand side'', i.e. corresponding to a physically reasonable 
matter source? As you will see in the subsequent sections,
the answer is ``yes'' provided that one further demands $h_{\mu \nu}$ 
to be the metric of an Einstein space of a $(D-1)$-dimensional 
Riemannian geometry. We call such a metric $g_{\mu\nu}$ a 
{\it G\"{o}del-type metric}.

In our first paper on this subject \cite{gks}, we have examined this
question in detail for the simple case $u_{k} = constant$. For the
choice of constant $u_{k}$, one finds that $u^{\mu}$ is a Killing vector,
thanks to the assumptions stated so far. In \cite{gks}, we showed 
that in all dimensions the Einstein equations are classically equivalent to
the field equations of general relativity with a charged dust
source provided that a simple $(D-1)$-dimensional Euclidean source-free 
Maxwell's equation is satisfied. Then the energy density of the
dust fluid is proportional to the Maxwell invariant $F^{2}$. We further 
demonstrated that the geodesics of the G{\"o}del-type metrics are
described by solutions of the $(D-1)$-dimensional Euclidean
Lorentz force equation for a charged particle. We also discussed the 
possible existence of examples of spacetimes containing closed timelike 
and closed null curves which violate causality and examples of spacetimes 
without any closed timelike or closed null curves where causality is
preserved. We showed that the G{\"o}del-type metrics with constant $u_{k}$
provide exact solutions to various kinds of supergravity theories
in five, six, eight, ten and eleven dimensions. All these exact 
solutions are fundamentally based on the vector field $u_{\mu}$ which 
satisfies the $(D-1)$-dimensional Maxwell's equation in the background 
of some $(D-1)$-dimensional Riemannian geometry with metric $h_{\mu\nu}$. 
It was in this respect that \cite{gks} gave not only a specific solution 
but in fact provided a whole class of exact solutions to each of the 
aforementioned theories. Moreover in \cite{gks}, explicit examples of 
exact solutions were separately constructed for the cases of both 
trivially flat and non-flat backgrounds $h_{\mu\nu}$, the latter of which 
included a conformally flat space, an Einstein space and a Riemannian 
Tangherlini solution in $D=4$. 

All of the solutions presented in \cite{gks} were mathematically simple and 
some of them were known beforehand; however, \cite{gks} certainly did 
provide a useful, nice and unified treatment of G{\"o}del-type solutions. 
In the present work, we follow on our promise to further consider the case 
when $u_{k} \neq constant$. 

We show that the information inherent in the Einstein equations calculated
from the conformally transformed G{\"o}del-type metric (or the G{\"o}del-type 
metric in a string frame) with a non-constant $u_{k}$ can be thought of as
classically equivalent to that of an Einstein-Maxwell-dilaton-3-form theory
for dimensions $D \geq 6$. The Maxwell and 3-form fields are related to the
vector field $u_{\mu}$ in a simple manner, whereas the dilaton field is
simply given by $\phi \equiv \ln{|u_{k}|}$. As a consequence of the special
construction employed in the formation of the 3-form field and the dilaton,
the field equations satisfied by these matter fields are shown to follow
from the ``Maxwell equation'' that the vector field $u_{\mu}$ satisfies. We
also comment on the possible choices of the $(D-1)$-dimensional Riemannian
spacetime backgrounds and derive the ``Maxwell'' and the ``dilaton'' 
equations explicitly in the background $h_{\mu \nu}$ using the ``Maxwell 
equation'' for $u_{\mu}$. We next show that the G{\"o}del-type metrics provide
exact solutions to the bosonic parts of various kinds of supergravity
theories in six, eight and ten dimensions. In this respect, rather than 
giving specific solutions, we provide a whole class of exact solutions
to each of these supergravity theories depending on the choice of their
respective backgrounds. As particular examples, we explicitly construct
solutions found by taking the $(D-1)$-dimensional flat Riemannian and 
Riemannian Tangherlini geometries as backgrounds. In the former case,
the solutions turn out to be the $D$-dimensional generalizations of the
well known Majumdar-Papapetrou metrics of $D=4$ dimensions. Since the $D=3$ 
case is special, we examine it separately and find a rich family of exact 
solutions then. Some of the long calculations and their technical details 
are presented in the two appendices following our conclusions.

\section{\label{emdil} Einstein-Maxwell-Dilaton-3-Form Theories}

It readily follows from the assumptions stated in the first paragraph of
the introduction that $h_{\mu \nu} \, u^{\nu}=0$ and the inverse
of the metric can be calculated as 
\begin{equation}
g^{\mu \nu}=\bar{h}^{\mu\nu}+(-1+\bar{h}^{\alpha\beta} \, u_{\alpha}\, 
u_{\beta}) \, u^{\mu} \, u^{\nu} 
+ u^{\mu} (\bar{h}^{\nu\alpha} \, u_{\alpha}\,) 
+ u^{\nu} (\bar{h}^{\mu\alpha} \, u_{\alpha}\,) \; . \label{invmet}
\end{equation}
Here $\bar{h}^{\mu\nu}$ is the $(D-1)$-dimensional inverse of $h_{\mu \nu}$;
i.e. \( \bar{h}^{\mu \nu} \, h_{\nu \alpha} = \bar{\delta}^{\mu}_{\;\alpha}\)
with $\bar{\delta}^{\mu}_{\;\alpha}$ denoting the $(D-1)$-dimensional 
Kronecker delta: \( \delta^{\mu}_{\;\alpha} = \bar{\delta}^{\mu}_{\;\alpha} + 
\delta^{\mu}_{\;k} \, \delta^{k}_{\;\alpha} \).

By defining the Christoffel symbols of $h_{\mu\nu}$ as
\begin{equation}
\gamma^{\mu}\,_{\alpha \beta}=\frac{1}{2}\, \bar{h}^{\mu\nu} \,
[h_{\nu\alpha, \, \beta} + h_{\nu\beta, \, \alpha} 
- h_{\alpha\beta, \, \nu}] \, ,
\label{gamchr}
\end{equation}
one finds that the Christoffel symbols of $g_{\mu\nu}$ are given by
\begin{equation}
\Gamma^{\mu}\,_{\alpha \beta} = \gamma^{\mu}\,_{\alpha \beta}
+ \frac{1}{2}\, (u_{\alpha} \, f^{\mu}\,_{\beta}
+ u_{\beta} \, f^{\mu}\,_{\alpha}) - \frac{1}{2} \, u^{\mu} \,
(u_{\alpha \vert \beta} + u_{\beta \vert \alpha}) \, . \label{chr}
\end{equation}
Here we use a vertical stroke to denote a covariant derivative with 
respect to $\gamma^{\mu}\,_{\alpha \beta}$ and 
\( f_{\alpha \beta} \equiv \partial_{\alpha} 
u_{\beta} - \partial_{\beta} u_{\alpha} \). We further assume that the 
indices of $u_{\mu}$ and $f_{\alpha\beta}$ are raised and lowered by 
the metric $g_{\mu\nu}$; thus e.g. 
\( f^{\mu}\,_{\nu}=g^{\mu \alpha}\, f_{\alpha \nu} \). Note that 
the definition of $f_{\alpha \beta}$ is left invariant if the ordinary 
derivatives are replaced by the covariant derivatives with respect to 
$\gamma^{\mu}\,_{\alpha \beta}$ (or $\Gamma^{\mu}\,_{\alpha \beta}$ for 
that matter). To remove any further ambiguity, we will denote a covariant 
derivative with respect to $\Gamma^{\mu}\,_{\alpha \beta}$ by 
$\nabla_{\mu}$ in what follows.

Now one can show that $u^{\mu}$ is not a Killing vector unlike the 
constant $u_{k}$ case examined in \cite{gks}. This follows from a 
number of identities that involve the vector $u^{\mu}$ and its covariant 
derivative $\nabla_{\beta} \, u_{\alpha}$ that are presented in detail in 
appendix \ref{appa}.

Let us now define $\phi \equiv \ln{|u_{k}|}$ and denote the energy 
momentum tensor of this scalar $\phi$ as 
\[ T_{\mu\nu}^{\phi} = (\nabla_{\mu} \phi) (\nabla_{\nu} \phi) 
-\frac{1}{2} \, g_{\mu\nu} \, (\nabla \, \phi)^{2} \, , \]
where \( (\nabla \, \phi)^{2} \equiv g^{\mu\nu} (\nabla_{\mu} \phi) 
(\nabla_{\nu} \phi) \). Similarly, one can also define
the energy momentum tensor for the ``Maxwell field'' $f_{\mu\nu}$ as
\[ T^{f}_{\mu \nu} \equiv f_{\mu \alpha} \, f_{\nu}\,^{\alpha} 
- \frac{1}{4} \, g_{\mu\nu} \, f^{2} \, , \]
where \( f^2 \equiv f^{\alpha \beta} \, f_{\alpha \beta} \). One can now
calculate the Einstein tensor to be
\begin{eqnarray}
G_{\mu\nu} & = & \hat{r}_{\mu\nu} - \frac{1}{2} \, g_{\mu\nu} \, \hat{r} 
- \frac{1}{2} \, T_{\mu\nu}^{\phi} + \frac{1}{2} \, 
T_{\mu\nu}^{f} - \nabla_{\mu} \nabla_{\nu} \phi 
+ \frac{1}{2} \, g_{\mu\nu} \, \Box \, \phi 
- \frac{1}{4} \, f^2 \, (u_{\mu} \, u_{\nu} + g_{\mu\nu}) \nonumber \\
& & + \frac{1}{2} \, u_{\mu} \, e^{\phi} \, \nabla_{\alpha} \, (e^{-\phi} \, 
f^{\alpha}\,_{\nu}) + \frac{1}{2} \, u_{\nu} \, e^{\phi} \, \nabla_{\alpha} 
\, (e^{-\phi} \, f^{\alpha}\,_{\mu}) - \frac{1}{2} \, g_{\mu\nu} \, 
\left( u_{\beta} \, e^{\phi} \, \nabla_{\alpha} \, 
(e^{-\phi} \, f^{\alpha\beta}) \right) ,
\label{emdein}
\end{eqnarray}
where \( \Box \, \phi \equiv g^{\mu\nu} \nabla_{\mu} \nabla_{\nu} 
\phi \). Here $\hat{r}_{\mu\nu}$ and $\hat{r}$ denote the Ricci tensor
and the Ricci scalar of $\gamma^{\mu}\,_{\alpha \beta}$, respectively.
[See appendix \ref{appa} for the details of how (\ref{emdein}) is obtained.]

One can already catch the glimpses of the field equations for the 
Einstein-Maxwell-dilaton theory at this stage. However the
$\nabla_{\mu} \nabla_{\nu} \phi$ term is obviously not very 
pleasing and needs to be discarded if one is to go for such an 
identification. This can be achieved by utilizing a conformal 
transformation on the metric $g_{\mu\nu}$. Let 
$\tilde{g}_{\mu\nu} = e^{2 \psi} \, g_{\mu\nu}$ 
(hence $\tilde{g}^{\mu\nu} = e^{-2 \psi} \, g^{\mu\nu}$), where $\psi$ 
is a smooth function. Using the results of \cite{wald}, one can calculate 
the Einstein tensor $\tilde{G}_{\mu\nu}$ associated with the metric 
$\tilde{g}_{\mu\nu}$ and its derivative operator $\tilde{\nabla}_{\mu}$
in terms of the quantities associated with $g_{\mu\nu}$ to be
\[ \tilde{G}_{\mu\nu} = G_{\mu\nu} - (D-2) \, \nabla_{\mu} \nabla_{\nu} \psi
+ (D-2) \, \psi_{\mu} \, \psi_{\nu} + (D-2) \, g_{\mu\nu} \, 
\left( \Box \, \psi + \frac{1}{2} \, (D-3) \, (\nabla \, \psi)^2 \right) \, .\]
Hence choosing the function $\psi$ as $\psi=\phi/(2-D)$, one can eliminate the 
$\nabla_{\mu} \nabla_{\nu} \phi$ term and, after a few simplifications, obtain
\begin{eqnarray}
\tilde{G}_{\mu\nu} & = & \frac{1}{2} \, u_{\mu} \, e^{\phi} \, 
\nabla_{\alpha} \, (e^{-\phi} \, 
f^{\alpha}\,_{\nu}) + \frac{1}{2} \, u_{\nu} \, e^{\phi} \, \nabla_{\alpha} 
\, (e^{-\phi} \, f^{\alpha}\,_{\mu}) - \frac{1}{2} \, g_{\mu\nu} \, 
\left( u_{\beta} \, e^{\phi} \, 
\nabla_{\alpha} \, (e^{-\phi} \, f^{\alpha\beta}) \right) \nonumber \\ 
& & + \hat{r}_{\mu\nu} - \frac{1}{2} \, g_{\mu\nu} \, \hat{r} 
+ \frac{1}{2} \, T_{\mu\nu}^{f} 
- \frac{1}{4} \, f^2 \, (u_{\mu} \, u_{\nu} + g_{\mu\nu}) \nonumber \\
& & - \frac{1}{2} \, g_{\mu\nu} \, \Box \, \phi 
+ \frac{3D-8}{4(D-2)} \, g_{\mu\nu} \, (\nabla \, \phi)^2
+ \frac{4-D}{2(D-2)} \, \phi_{\mu} \, \phi_{\nu} \, .
\label{emdeintil}
\end{eqnarray}

Now one has to transform the right hand side of (\ref{emdeintil}) in such
a way that all quantities and especially covariant derivatives are only
related to the conformally transformed metric (or the string metric)
$\tilde{g}_{\mu\nu}$. The details of this calculation are not very
illuminating and so we present them in appendix \ref{appb}. The result
is the following:
\begin{eqnarray}
\tilde{G}_{\mu\nu} & = & \hat{r}_{\mu\nu} 
+ \frac{4-D}{2(D-2)} \, \tilde{T}_{\mu\nu}^{\phi} 
+ \frac{1}{2} \, e^{\frac{2 \phi}{2-D}} \, \tilde{T}_{\mu\nu}^{f} \nonumber \\
& & - \frac{1}{2} \, \tilde{g}_{\mu\nu} 
\left( e^{- \frac{2 \phi}{2-D}} \, \hat{r} 
+ \widetilde{\Box} \, \phi + \frac{1}{2} \, e^{\frac{2 \phi}{2-D}} \, 
\tilde{f}^2 + \tilde{u}_{\beta} \, \tilde{\nabla}_{\alpha} \,
(e^{\frac{2 \phi}{2-D}} \, \tilde{f}^{\alpha\beta}) \right) \nonumber \\
& & + \frac{1}{2} \tilde{u}_{\mu} \left( \tilde{\nabla}_{\alpha} 
(e^{\frac{2 \phi}{2-D}} \tilde{f}^{\alpha}\,_{\nu}) - \frac{1}{4}  
e^{\frac{4 \phi}{2-D}} \tilde{f}^2 \tilde{u}_{\nu} \right)
+ \frac{1}{2} \tilde{u}_{\nu} \left( \tilde{\nabla}_{\alpha} 
(e^{\frac{2 \phi}{2-D}} \tilde{f}^{\alpha}\,_{\mu}) - \frac{1}{4}  
e^{\frac{4 \phi}{2-D}} \tilde{f}^2 \tilde{u}_{\mu} \right) .
\label{emdeinst}
\end{eqnarray}
We would like to emphasize in passing that here we take 
$\tilde{u}_{\mu} = u_{\mu}$, $\tilde{f}_{\mu\nu} = f_{\mu\nu}$
and the indices on these quantities are raised using $\tilde{g}^{\mu\nu}$
(see appendix \ref{appb} for details).

Let us now define a 3-form field $\tilde{H}_{\mu\tau\sigma}$ as
\begin{equation}
\tilde{H}_{\mu\tau\sigma} \equiv \tilde{f}_{\mu\tau} \, \tilde{u}_{\sigma}
+ \tilde{f}_{\tau\sigma} \, \tilde{u}_{\mu} +
\tilde{f}_{\sigma\mu} \, \tilde{u}_{\tau} \, . \label{Hdef}
\end{equation}
If we denote the energy momentum tensor for this 3-form field
$\tilde{H}_{\mu\tau\sigma}$ as
\[ \tilde{T}_{\mu\nu}^{H} \equiv \tilde{H}_{\mu\tau\sigma} \,
\tilde{H}_{\nu}\,^{\tau\sigma} - \frac{1}{6} \, \tilde{g}_{\mu\nu} 
\, \tilde{H}^{2} \, , \]
where \( \tilde{H}^{2} \equiv \tilde{H}_{\mu\tau\sigma} 
\, \tilde{H}^{\mu\tau\sigma} \, ,\) one finds by the help of
\begin{eqnarray}
\tilde{H}_{\mu\tau\sigma} \, \tilde{H}_{\nu}\,^{\tau\sigma} & = &
-2 \, e^{- \frac{2 \phi}{2-D}} \, \tilde{f}_{\mu\tau} \, 
\tilde{f}_{\nu}\,^{\tau} + \tilde{f}^{2} \, \tilde{u}_{\mu} \, \tilde{u}_{\nu} 
-2 \, e^{- \frac{4 \phi}{2-D}} \, (\tilde{\nabla}_{\mu} \phi) \, 
(\tilde{\nabla}_{\nu} \phi) \nonumber \\
& & -2 \, e^{- \frac{2 \phi}{2-D}} \, (\tilde{\nabla}_{\tau} \phi) \,  
(\tilde{f}_{\nu}\,^{\tau} \, \tilde{u}_{\mu} + 
\tilde{f}_{\mu}\,^{\tau} \, \tilde{u}_{\nu}) \, , \label{HmuHnu} \\ 
\tilde{H}^{2} & = & -3 \, e^{- \frac{2 \phi}{2-D}} \, \tilde{f}^{2} 
-6 \, e^{- \frac{4 \phi}{2-D}} \, (\tilde{\nabla} \phi)^{2} \, , \label{Hkare}
\end{eqnarray}
that (\ref{emdeinst}) can be written as
\begin{equation} 
\tilde{G}_{\mu\nu} = \hat{r}_{\mu\nu} 
- \frac{1}{2} \, \tilde{g}_{\mu\nu} \, e^{- \frac{2 \phi}{2-D}} \, \hat{r}
+ \frac{1}{4} \, e^{\frac{4 \phi}{2-D}} \, \tilde{T}_{\mu\nu}^{H}
+ e^{\frac{2 \phi}{2-D}} \, \tilde{T}_{\mu\nu}^{f} 
+ \frac{1}{D-2} \, \tilde{T}_{\mu\nu}^{\phi} \, ,
\label{einson}
\end{equation}
provided that
\begin{eqnarray}
\tilde{\nabla}_{\mu} (e^{\frac{2 \phi}{2-D}} \, \tilde{f}^{\mu\nu})
& = & \frac{1}{2} \, e^{\frac{4 \phi}{2-D}} \, \tilde{H}^{\nu\tau\sigma} \,
\tilde{f}_{\tau\sigma} \, , \label{maxson} \\
\widetilde{\Box} \, \phi + \frac{1}{2} \, e^{\frac{2 \phi}{2-D}} 
\, \tilde{f}^{2} + \frac{1}{6} \, e^{\frac{4 \phi}{2-D}} \, \tilde{H}^{2} 
& = & 0 \, , \label{dilson} \\
\tilde{\nabla}_{\mu} (e^{\frac{4 \phi}{2-D}} \, \tilde{H}^{\mu\nu\alpha})
& = & 0 \, . \label{Hson}
\end{eqnarray}

Before making an interpretation of these equations, let us in fact observe
that both (\ref{dilson}) and (\ref{Hson}) follow from (\ref{maxson}). This
is seen by noting that (\ref{maxson}) can be equivalently written as
\begin{equation}
\tilde{\nabla}_{\mu} \tilde{f}^{\mu}\,_{\nu} = \frac{1}{2} \, 
e^{\frac{2 \phi}{2-D}} \, \tilde{f}^{2} \, \tilde{u}_{\nu} -
\frac{D}{2-D} \, (\tilde{\nabla}_{\sigma} \phi) \, 
\tilde{f}^{\sigma}\,_{\nu} \, . \label{otuz}
\end{equation}
Contracting this equation by $\tilde{u}^{\nu}$ and using the identities
\[ \tilde{u}_{\mu} \, \tilde{u}^{\mu} = - e^{- \frac{2 \phi}{2-D}} 
\quad \quad \mbox{and} \quad \quad 
\tilde{u}^{\mu} \, \tilde{f}_{\mu\nu} = e^{- \frac{2 \phi}{2-D}} \,
\tilde{\nabla}_{\nu} \, \phi \, , \]
one obtains
\begin{equation}
\widetilde{\Box} \, \phi = ( \tilde{\nabla} \, \phi )^2 \label{otuzbir}
\end{equation}
which is itself equivalent to (\ref{dilson}) by (\ref{Hkare}).
Similarly using 
\[ \tilde{H}^{\nu\tau\sigma} \, \tilde{f}_{\tau\sigma} = 
\tilde{f}^{2} \, \tilde{u}^{\nu} - 2 \, (\tilde{\nabla}_{\sigma} \, \phi) \,
e^{- \frac{2 \phi}{2-D}} \, \tilde{f}^{\nu\sigma} \, , \]
in (\ref{maxson}), contracting this new equation by $\delta^{\alpha}_{\;k}$ 
(which is equal to $-u_{k} \, e^{\frac{2 \phi}{2-D}} \, \tilde{u}^{\alpha}$), 
adding the copies of the resulting equation that are obtained by taking 
the cyclic permutations of the three indices, one finds (\ref{Hson}). 

In fact using $K^{\sigma}\,_{\mu\nu}$ introduced in appendix \ref{appb}, 
(\ref{maxson}) (or equivalently (\ref{otuz})) can be transformed into
\[ \nabla_{\mu} f^{\mu}\,_{\nu} = \frac{1}{2} \, f^{2} \, u_{\nu} 
+ 2 \, \phi_{\sigma} \, f^{\sigma}\,_{\nu} \]
and by using (\ref{invmet}) and (\ref{chr}),
this can be further written as
\begin{equation}
(\bar{h}^{\mu\alpha} \, f_{\alpha\nu})_{\vert \mu} + 
(\bar{h}^{\mu\sigma} \, u_{\sigma} \, \phi_{\nu})_{\vert \mu} =
\bar{h}^{\mu\alpha} \, f_{\alpha\nu} \, \phi_{\mu} +
\bar{h}^{\mu\sigma} \, \phi_{\mu} \, u_{\sigma} \, \phi_{\nu} +
\bar{h}^{\mu\sigma} \, \phi_{\sigma} \, u_{\nu \vert \mu} 
\label{maxinh}
\end{equation}
in the background $h_{\mu\nu}$. Similarly (\ref{dilson}) (or 
equivalently (\ref{otuzbir})) is simply given by
\begin{equation}
\Box_{h} \, e^{-\phi} \equiv \frac{1}{\sqrt{|h|}} \, \partial_{\mu}
(\sqrt{|h|} \, \bar{h}^{\mu\nu} \, \partial_{\nu} \, e^{-\phi}) = 0
\label{dilinh}
\end{equation}
in terms of the background $h_{\mu\nu}$. The information inherent in
(\ref{dilinh}) is naturally contained in (\ref{maxinh}), of course.

In retrospect we have shown that the conformally transformed metric
(or the string metric)
\[ \tilde{g}_{\mu\nu} = e^{\frac{2 \phi}{2-D}} \, 
(h_{\mu \nu}-u_{\mu} u_{\nu}) \, , \] 
our choice of $u_{\mu}$ (and hence $f_{\mu\nu}$) and our
construction of the 3-form field $\tilde{H}_{\mu\tau\sigma}$ solve
the Einstein-Maxwell-dilaton-3-form field equations (\ref{einson}),
(\ref{maxson}), (\ref{dilson}), (\ref{Hson}) in $D$-dimensions provided 
the background $h_{\mu\nu}$ is chosen suitably so that the contribution 
of the first two terms on the right hand side of (\ref{einson}) can 
be controlled and given a physically reasonable matter interpretation 
and the Maxwell equation (\ref{maxinh}) (and hence the dilaton equation 
(\ref{dilinh})) in the background $h_{\mu\nu}$ are satisfied. 

We should perhaps emphasize that the very forms of (\ref{maxson}), 
(\ref{dilson}) and (\ref{Hson}) all follow naturally from the way the
Maxwell field and the 3-form field are constructed; the coefficients 
that show up in these equations are not in any way determined by a 
further requirement or a framework such as supersymmetry.

Now let us recall that by assumption $u^{\mu}$ is taken as a timelike 
vector and for the string metric $\tilde{g}_{\mu\nu}$ to have the 
correct Minkowskian signature, the background ${h}_{\mu\nu}$ has to 
necessarily be taken as the metric of a $(D-1)$-dimensional Riemannian
spacetime. However note that a crucial requirement here is to be able
to choose the background ${h}_{\mu\nu}$ so that the first two terms 
on the right hand side of (\ref{einson}) are kept under control. This 
requirement indeed constrains the choice of allowable ${h}_{\mu\nu}$ 
further. A few of the acceptable backgrounds that first come in to mind 
are as follows: flat Riemannian spacetimes, flat Riemannian Tangherlini 
solutions, flat Riemannian Myers-Perry solutions, Ricci-flat K\"{a}hler 
manifolds (provided that $D-1$ is even, of course) and suitably chosen 
gravitational instanton solutions. Depending on whether one is willing 
to accommodate an extra perfect fluid source on the right hand side of 
(\ref{einson}), one can also try out the de Sitter versions of some of 
the aforementioned spacetimes.

We should also make a few remarks regarding the similarity between our
G{\"o}del-type metrics with the metrics used in Kaluza-Klein reductions in
string theory. In a typical Kaluza-Klein reduction from $D$-dimensions
to $(D-1)$-dimensions, the process is usually carried on a spatial dimension
and the theory obtained in the lower dimension has the usual Minkowskian
signature just like its parent theory. Here, contrary to what is done in
the Kaluza-Klein mechanism, the G{\"o}del-type metrics are used in obtaining
a $D$-dimensional theory starting from a $(D-1)$-dimensional one, and this
is done by the use of backgrounds $h_{\mu\nu}$ with Euclidean signature. If
anything, such backgrounds could only emerge after a Kaluza-Klein reduction
performed on the time direction of a usual Minkowskian signature theory,
and hence as objects living in Euclidean signature gravity (or perhaps
supergravity) theories. (In this respect, \cite{cllp} is a pioneering
work that studies how Euclidean signature supergravities arise by 
compactifying $D=11$ supergravity or type IIB supergravity on a torus that
includes the time direction.)

\section{\label{vardims} Solutions in Various Dimensions}

Let us now look for simple theories whose field equations contain the
Einstein-Maxwell-dilaton-3-form field equations (\ref{einson}),
(\ref{maxson}), (\ref{dilson}), (\ref{Hson}) that we have derived in the
previous section. The presence of a 3-form itself suggests that one
should turn to dimensions $D \ge 6$. As will become apparent in what
follows, one can use any one of the backgrounds listed in the second
last paragraph of section \ref{emdil} to construct solutions to the 
bosonic parts of some supergravity theories in six, eight and ten dimensions. 
We do not claim that these are the only possible theories for which the 
techniques we present are applicable, the theories we consider here are merely 
examples. We would also like to stress out that our aim here is just to 
show how one can use the construction presented in the previous section 
to obtain solutions to these theories. A detailed analysis of these 
solutions and a consideration of how much supersymmetry, if any, they 
preserve is far beyond the scope of this work and, apart from mentioning 
possible directions to look for, here we completely refrain from delving 
into such delicate issues. 

\subsection{\label{6dim} Six dimensions}

In our conventions, the bosonic part of the gauged $D=6, N=2$ supergravity 
\cite{sez} reduces to the following field equations when all the
scalars of the hypermatter $\phi^{a}$ and the 2-form field $B_{\mu \nu}$
in the theory are set to zero and the dilaton $\varphi$ is taken to be
non-constant \footnote{The constant factor in front of the very last term 
of (4.17) of \cite{sez} is incorrect. Here we fix this and use the correct 
one.}:
\begin{eqnarray}
R_{\mu \nu} & = & 2 \, e^{\sqrt{2} \varphi} 
\, F_{\mu \rho} \, F_{\nu}\,^{\rho} 
+ e^{2 \sqrt{2} \varphi} \, G_{\mu \rho \sigma} \, G_{\nu}\,^{\rho \sigma} 
+ 2 \, (\nabla_{\mu} \varphi) \, (\nabla_{\nu} \varphi)
-\frac{1}{\sqrt{2}} \, g_{\mu\nu} \, \Box \varphi , \label{d=6eq1}\\
\nabla_{\mu} ( e^{\sqrt{2} \varphi} \, F^{\mu \nu}) & = &
e^{2 \sqrt{2} \varphi} \, \, G^{\nu \rho \sigma} 
\, F_{\rho \sigma} \, , \label{d=6eq2}\\
\nabla_{\mu} ( e^{2 \sqrt{2} \varphi} \, G^{\mu \nu \rho}) 
& = & 0 \, , \label{d=6eq3}\\
\Box \varphi & = & \frac{1}{3 \sqrt{2}} \, e^{2 \sqrt{2} \varphi} 
\, G_{\mu \nu \rho} \, G^{\mu \nu \rho} + \frac{1}{2 \sqrt{2}} \, 
e^{\sqrt{2} \varphi} \,F_{\mu \nu} \, F^{\mu \nu} \, . \label{d=6eq4}
\end{eqnarray}

Here all Greek indices run from 0 to 5 and $G_{\mu \nu \rho}$ is given by
\begin{equation}
G_{\mu \nu \rho} = F_{\mu \nu} \, A_{\rho} + F_{\nu \rho} \, A_{\mu} 
+ F_{\rho \mu}\, A_{\nu} \, \label{gdef}
\end{equation}
and instead of a Yang-Mills field, we have taken an ordinary vector 
field $A_{\mu}$ to be present. In fact using (\ref{d=6eq4}) in (\ref{d=6eq1}),
one finds that the Einstein tensor satisfies
\begin{equation}
G_{\mu \nu} = 2 \, e^{\sqrt{2} \varphi} \, T_{\mu\nu}^{F} +
e^{2 \sqrt{2} \varphi} \, T_{\mu\nu}^{G} + 2 \, T_{\mu\nu}^{\varphi}
\label{d=6eq5}
\end{equation}
for this theory. Notice the striking resemblance of this theory with 
the model we described in Section \ref{emdil}. 

Now let $h_{\mu\nu}$ be any Ricci-flat Riemannian metric in five dimensions so
that $\hat{r}_{\mu\nu}=\hat{r}=0$ and 
\begin{equation}
g_{\mu \nu}= 2 \, e^{\sqrt{2} \varphi} \, (h_{\mu \nu}-u_{\mu}\, u_{\nu}) \, .
\label{d=6met}
\end{equation}
Moreover take $A_{\mu}=u_{\mu}$ so that $F_{\mu\nu}=f_{\mu\nu}$ and
$G_{\mu\nu\rho}=\tilde{H}_{\mu\nu\rho}$ as in Section \ref{emdil}. Then a
careful comparison of (\ref{d=6eq5}), (\ref{d=6eq2}), (\ref{d=6eq3}),
(\ref{d=6eq4}) with (\ref{einson}), (\ref{maxson}), (\ref{Hson}) and
(\ref{dilson}), respectively, shows that these two sets of equations are
identical provided $\phi=-2 (\sqrt{2} \varphi + \ln{2})$.

Hence the conformally transformed G{\"o}del-type metric (\ref{d=6met})
becomes an exact solution of $D=6, N=2$ supergravity theory provided
$h_{\mu\nu}$ is chosen as any Ricci-flat Riemannian metric in five dimensions
and $u_{\mu}$ satisfies (\ref{maxinh}) in this background. As stated earlier,
our aim here is just to show how the techniques of section \ref{emdil} can
be used to find solutions. A comparison of the solutions that can be
obtained here with the (supersymmetric) ones given in \cite{guto} and
\cite{cari} (and the references therein) and an investigation of how much
supersymmetry they preserve is certainly worth further study.

\subsection{\label{8dim} Eight dimensions}

The bosonic part of the gauged $D=8, N=1$ supergravity theory coupled to
$n$ vector multiplets \cite{sals} has field equations which are very
similar to the field equations of the gauged $D=6$, $N=2$ supergravity 
theory that we have examined in subsection \ref{6dim}. Taking an ordinary 
vector field instead of a Yang-Mills field and setting the 2-form field 
$B_{MN}$ equal to zero, one similarly has
\begin{equation}
G_{MNP} = F_{MN} \, A_{P} + F_{NP} \, A_{M} + F_{PM}\, A_{N} \, , \label{g8def}
\end{equation}
where now capital Latin indices run from 0 to 7. We also set all the 
scalars in the theory to zero apart from the dilaton $\sigma$ which we
take as non-constant. These assumptions lead to the following field 
equations (see (26) of \cite{sals})
\begin{eqnarray}
R_{MN} & = & 2 e^{\sigma} \, F_{MP} \, F_{N}\,^{P} 
+ e^{2 \sigma} \, G_{MPS} \, G_{N}\,^{PS} + \frac{3}{2} \,
(\nabla_{M} \sigma) \, (\nabla_{N} \sigma) - \frac{1}{2} \,
g_{MN} \, \Box \sigma , \label{d=8eq1}\\
\nabla_{M} (e^{\sigma} \, F^{MN}) & = & e^{2 \sigma} 
\, G^{NPS} \, F_{PS} \, , \label{d=8eq2}\\
\nabla_{M} (e^{2 \sigma} \, G^{MNP}) & = & 0 \, , \label{d=8eq3}\\
\Box \sigma & = & \frac{2}{9} \, e^{2 \sigma} \, G_{MNP} \, G^{MNP}
+ \frac{1}{3} \, e^{\sigma} \, F_{MN} \, F^{MN} \, , \label{d=8eq4}
\end{eqnarray}
in our conventions. In fact using (\ref{d=8eq4}) in (\ref{d=8eq1}),
one finds that the Einstein tensor satisfies
\begin{equation}
G_{MN} =  2 \, e^{\sigma} \, T_{MN}^{F} +
e^{2 \sigma} \, T_{MN}^{G} + \frac{3}{2} \, T_{MN}^{\sigma} \, .
\label{d=8eq5}
\end{equation}

Now let $h_{MN}$ be any Ricci-flat Riemannian metric in seven dimensions so
that $\hat{r}_{MN}=0$, $\hat{r}=0$ and 
\begin{equation}
g_{MN}= 2 \, e^{\sigma} \, (h_{MN}-u_{M}\, u_{N}) \, .
\label{d=8met}
\end{equation}
Moreover take $A_{M}=\tilde{u}_{M}$ so that $F_{MN}=\tilde{f}_{MN}$ and
$G_{MNP}=\tilde{H}_{MNP}$ similarly to what we did in subsection \ref{6dim}.
Then a careful comparison of (\ref{d=8eq5}), (\ref{d=8eq2}), (\ref{d=8eq3}),
(\ref{d=8eq4}) with (\ref{einson}), (\ref{maxson}), (\ref{Hson}) and
(\ref{dilson}), respectively, shows that these two sets of equations are
identical provided $\phi=-3(\sigma + \ln{2})$.

One again arrives at the conclusion that the conformally transformed 
G{\"o}del-type metric (\ref{d=8met}) yields an exact solution to the
gauged $D=8, N=1$ supergravity with matter couplings provided
$h_{MN}$ is chosen as any Ricci-flat Riemannian metric in seven dimensions
and $u_{M}$ satisfies (\ref{maxinh}) in this background. The conditions
on $u_{M}$ under which these solutions are supersymmetric should be 
examined further.

\subsection{\label{10dim} Ten dimensions}

In our conventions, the following field equations belong to type IIB
supergravity with only a graviton, a dilaton and a 3-form gauge field
present \cite{schw}:
\begin{eqnarray}
R_{MN} & = & \frac{1}{2} \, (\nabla_{M} \sigma) \, (\nabla_{N} \sigma)
+ \frac{1}{4} \, e^{\sigma} \, F_{MPQ} \, F_{N}\,^{PQ}
- \frac{1}{48} \, e^{\sigma} \, g_{MN} \, F_{PQR} \, F^{PQR} \, ,
\label{d=10eq1} \\
\nabla_{M} (e^{\sigma} \, F^{MNP}) & = & 0 \, , \label{d=10eq2}\\
\Box \sigma & = & \frac{1}{12} \, e^{\sigma} \, F_{MNP} \, F^{MNP} \, .
\label{d=10eq3}
\end{eqnarray}
Here all Latin indices run from 0 to 9. Using (\ref{d=10eq1}) and 
(\ref{d=10eq3}), the Einstein tensor can be shown to satisfy
\begin{equation}
G_{MN} =  \frac{1}{4} \, e^{\sigma} \, T_{MN}^{F} +
\frac{1}{2} \, T_{MN}^{\sigma} \, .
\label{d=10eq4}
\end{equation}

Comparing (\ref{d=10eq4}), (\ref{d=10eq2}), (\ref{d=10eq3})
with (\ref{einson}), (\ref{Hson}) and (\ref{dilson})
(with the 2-form field $\tilde{f}$ set identically equal to zero 
in the latter set), respectively, these two sets of equations can be
shown to be identical provided $\tilde{H}=F$ and $\phi=-2 \sigma$.

Therefore, provided $h_{MN}$ is chosen as any Ricci-flat Riemannian metric 
in nine dimensions and $u_{M}$ satisfies (\ref{maxinh}) in this 
background, a 2-form field $f_{MN}$ can be formed in the usual way
which in turn can be used in constructing a 3-form field
\[ F_{MNP} = f_{MN} \, u_{P} + f_{NP} \, u_{M} + f_{PM}\, u_{N} \]
that can be used together with the conformally transformed 
G{\"o}del-type metric 
\[ g_{MN}= e^{\sigma/2} \, (h_{MN}-u_{M}\, u_{N}) \]
to solve (\ref{d=10eq1}), (\ref{d=10eq2}) and (\ref{d=10eq3}).

\section{\label{solsec} Special Solutions Constructed by Simple Choices
of Backgrounds}

Just to give an idea as to how one goes about finding actual solutions, we 
will now consider as simple examples the solutions that can be constructed 
by taking the $(D-1)$-dimensional flat Riemannian and Riemannian Tangherlini 
geometries as the backgrounds. These can, of course, be considered as 
solutions of the supergravity theories examined in section \ref{vardims}
above.

\subsection{\label{ssflat} Spacetimes with $(D-1)$-dimensional flat Riemannian 
solutions as backgrounds}

Obviously, the simplest possible background one can think of is the 
$(D-1)$-dimensional flat Riemannian geometry. Now without loss of generality,
let us assume that the special fixed coordinate has been chosen so that
$k=0$. Thus our background reads
\[ ds_{D-1}^{2} = (dx^{1})^{2} + (dx^{2})^{2} + \cdots 
+ (dx^{D-1})^{2} = \bar{\delta}_{ij} \, dx^{i} \, dx^{j} \, , \]
where, as in a standard spacetime decomposition, Latin indices $i,j$ range
from 1 to $D-1$. While we are at it, let us furthermore assume that $u_{\mu}$
has only one non-zero component and so 
\( u_{\mu} = e^{\phi} \, \delta_{\mu}^{0} , \)
where $\phi$ depends on all $x^{\alpha}$ except for $x^{0}$. It immediately
follows that the only non-vanishing components of $f_{\mu\nu}$ are given
by \( f_{i0}=- f_{0i}= \phi_{i} \, e^{\phi} . \)

Now it is very easy to see that (\ref{dilinh}) yields 
\( \partial^{2} \, e^{-\phi} = 0 \, , \)
where 
\( \partial^{2} \equiv \bar{\delta}^{ij} \, \partial_{i} \, \partial_{j} \)
is the usual Laplacian operator of Euclidean geometry. One can now also
show that (\ref{maxinh}) is identically satisfied. Thus, in this simplest
example, the $D$-dimensional line element is found as
\begin{equation} 
ds^{2} = e^{\frac{2 \phi}{2-D}} \left[ (dx^{1})^{2} + \cdots 
+ (dx^{D-1})^{2} - e^{2 \phi} \, (dx^{0})^{2} \right] \, , \label{majp}
\end{equation}
where $e^{-\phi}$ is any harmonic function that solves 
\( \partial^{2} \, e^{-\phi} = 0 . \)

Actually the G{\"o}del-type metric (\ref{majp}) is the $D$-dimensional
generalization of the well known Majumdar-Papapetrou \cite{maj,pap} metric
of $D=4$ dimensions, and has recently been used in constructing multi shell
models for removing the multi black hole singularities of such spacetimes
\cite{gurbur}. We should emphasize that the G{\"o}del-type metric (\ref{majp})
is a solution to the supergravity models considered in section \ref{vardims}.
Specifically, we find the following:

{\bf i)} $D=6$ model described in subsection \ref{6dim}:

(\ref{majp}) now reduces to
\[ ds^{2} = e^{-\phi/2} \left[ (dx^{1})^{2} + \cdots 
+ (dx^{5})^{2} \right] - e^{3 \phi/2} \, (dx^{0})^{2} \, . \]
As it has been carefully described in subsection \ref{6dim}, in this case
the field content is properly covered by taking 
\( A_{\mu} = u_{\mu} = e^{\phi} \, \delta_{\mu}^{0} \) and hence 
\( F_{\mu\nu} = f_{\mu\nu} \) and \( G_{\mu\nu\rho} = \tilde{H}_{\mu\nu\rho} \)
as in Section \ref{emdil}. The harmonic function $e^{-\phi}$ can be taken as
\[ e^{-\phi} = 1 + \sum_{k=1}^{N} \, \frac{m_{k}}{|\vec{r}-\vec{a}_{k}|^3} \]
in analogy with the Majumdar-Papapetrou construction and this can be used
to find the dilaton field $\varphi$ by the relation 
\( \phi = -2 (\sqrt{2} \varphi + \ln{2}) \) (as described in subsection 
\ref{6dim}). Here the parameters $m_{k}$ can be thought of as denoting the
masses of the $N$ point particles located at \( \vec{r}=\vec{a}_{k} \) each
(see \cite{gurbur} for details).

{\bf ii)} $D=8$ model described in subsection \ref{8dim}:

A similar analysis as done for $D=6$ above can also be carried out for
$D=8$ by following the discussion at subsection \ref{8dim}. One finds
that this time (\ref{majp}) reduces to
\[ ds^{2} = e^{-\phi/3} \left[ (dx^{1})^{2} + \cdots 
+ (dx^{7})^{2} \right] - e^{5 \phi/3} \, (dx^{0})^{2} \, , \]
and taking \( A_{M} = \tilde{u}_{M} = e^{\phi} \, \delta_{\mu}^{0} \)
(and hence $F_{MN}=\tilde{f}_{MN}$ and $G_{MNP}=\tilde{H}_{MNP}$), the
model described in subsection \ref{8dim} is solved provided the
harmonic function $e^{-\phi}$ is now taken as
\[ e^{-\phi} = 1 + \sum_{k=1}^{N} \, \frac{m_{k}}{|\vec{r}-\vec{a}_{k}|^5} \]
with a similar analogy and notation as before. The dilaton field $\sigma$
is found by using the relation $\phi=-3(\sigma + \ln{2})$ in this case.

{\bf iii)} $D=10$ model described in subsection \ref{10dim}:

Finally the $D=10$ theory described in subsection \ref{10dim} can be
covered in a similar manner. The metric
\[ ds^{2} = e^{-\phi/4} \left[ (dx^{1})^{2} + \cdots 
+ (dx^{9})^{2} \right] - e^{7 \phi/4} \, (dx^{0})^{2} \, , \]
with the properly constructed 3-form field $F_{MNP}=\tilde{H}_{MNP}$
(as explained in subsection \ref{10dim}) and the
harmonic function $e^{-\phi}$ 
\[ e^{-\phi} = 1+\sum_{k=1}^{N} \, \frac{m_{k}}{|\vec{r}-\vec{a}_{k}|^7} \, ,\]
defined in analogy with the previous cases, make up the field content of
the $D=10$ model of subsection \ref{10dim}. The dilaton field $\sigma$
in this case is simply given by $\phi = - 2 \sigma$.

Note that the ansatz we used for $u_{\mu}$ was rather simple. One can, of 
course, use more complicated ones such as \( u_{\mu}= u_{i} \, \delta_{\mu}^{i}
+ e^{\phi} \, \delta_{\mu}^{0} \, , \) with the functions $u_{i}$ and $\phi$
depending on any number of $x^{j}$ but $x^{0}$, which will cause some form 
of rotation in the spacetime described by the metric $\tilde{g}_{\mu\nu}$.
It is clear that there is a vast number of possibilities that can be tried
out. 

\subsection{\label{sstang} Spacetimes with $(D-1)$-dimensional Riemannian 
Tangherlini solutions as backgrounds}

Consider the line element corresponding to the $(D-1)$-dimensional Riemannian
Tangherlini solution
\begin{equation} 
ds_{D-1}^{2} = \zeta(r) \, dt^2 + \frac{dr^2}{\zeta(r)} + r^2 
\, d\Omega_{D-3}^{2} \, , \label{tangback}
\end{equation}
where
\[ \zeta(r) = 1- 2 \, m \, r^{4-D} \, , \;\;\;\; (D \ge 4) \, , \]
$m>0$ is the constant mass parameter, $d\Omega_{D-3}^{2}$ is the metric 
on the $(D-3)$-dimensional unit sphere and $r$ is the usual radial coordinate
that defines this sphere \cite{gks,gs3}. [Even though the background is 
flat when $D=4$, we keep it for the discussion that will follow.]

Let the special fixed coordinate $x^{k}$ be $x^{D-1}$ and assume that
\( u_{\mu}=u(r) \, \delta_{\mu}^{0} + e^{\phi(r)} \, \delta_{\mu}^{D-1} \, . \)
Then \( f_{\mu\nu} = (\delta_{\mu}^{r} \, \delta_{\nu}^{0} - 
\delta_{\mu}^{0} \, \delta_{\nu}^{r}) \, u^{\prime} +
(\delta_{\mu}^{r} \, \delta_{\nu}^{D-1} - 
\delta_{\mu}^{D-1} \, \delta_{\nu}^{r}) \, \phi^{\prime}
\, e^{\phi} \, , \)
where prime denotes derivative with respect to $r$. Now (\ref{dilinh})
implies that
\begin{equation} 
(r^{D-3} \, \zeta \, (e^{-\phi})^{\prime})^{\prime} = 0 \, , 
\label{phieqn}
\end{equation}
which is easily integrated as
\begin{eqnarray}
e^{-\phi} & = & a + \left \{\begin{array}{ll} \frac{b}{2m (D-4)} 
\, \ln{\zeta} \, , & (D \ge 5) \\
\frac{b}{1-2m} \, \ln{r} \, , & (D=4) \end{array}
\right. \, , \label{phisol}
\end{eqnarray}
for some real integration constants $a$ and $b$. One can now calculate the
Christoffel symbols $\gamma^{\mu}\,_{\alpha\beta}$ of the background
and use these in (\ref{maxinh}) to see that the only nontrivial 
component of (\ref{maxinh}) (independent of the information provided by
(\ref{phieqn})) is obtained when $\nu=0$ and in that case $u(r)$
satisfies
\begin{eqnarray}
u^{\prime\prime} + \left( \frac{D-3}{r} - 2 \, \phi^{\prime} \right) \,
u^{\prime} + \left( \frac{\phi^{\prime} \, \zeta^{\prime}}{2 \, \zeta} 
\right) \, u & = & 0 \, , \;\;\;\; (D \ge 5) \, , \label{ueqn} \\
u^{\prime\prime} + \left( \frac{1}{r} - 2 \, \phi^{\prime} \right) \,
u^{\prime} & = & 0 \, , \;\;\;\; (D = 4) \, . \label{d=4ueqn}
\end{eqnarray}
Unfortunately, for $\zeta(r)$ and $\phi(r)$ given as above, one can not
find an explicit solution to the second order linear ordinary differential
equation (\ref{ueqn}) in terms of known functions. However, one can show 
implicitly that a physically acceptable solution to (\ref{ueqn}) exists 
and a numerical solution can be constructed by starting at 
$r \rightarrow \infty$ and coming in toward the outmost singularity 
at some $r_{s}>0$. Fortunately, things are a little better for $D=4$ 
since (\ref{d=4ueqn}) can be integrated to give
\[ u(r) = c_{1} - \frac{c_{2}}{b} \, (1-2m) \, e^{\phi(r)} =
c_{1} - \frac{c_{2} \, (1-2m)^{2}}{b (a(1-2m) + b \ln{r})} \, , 
\;\;\;\; (D=4) \, , \] 
for some real integration constants $c_{1}$ and $c_{2}$.

The $D$-dimensional line element reads
\begin{eqnarray}
ds^2 & = & \left \{\begin{array}{ll}
e^{\frac{2 \phi}{2-D}} \left( \zeta(r) dt^2 + \frac{dr^2}{\zeta(r)} 
+ r^2 d\Omega_{D-3}^{2} - (u(r) dt + e^{\phi(r)} dx^{D-1})^2 
\right), & (D \ge 5), \\
\frac{a(1-2m) + b \ln{r}}{1-2m} \left( (1-2m) dt^2 
+ \frac{dr^2}{1-2m} + r^2 d\theta^{2} \right. \\
\qquad \qquad \quad \left. 
-\left( (c_{1} - \frac{c_{2} (1-2m)^{2}}{b (a(1-2m) + b \ln{r})})dt 
+ \frac{1-2m}{a(1-2m) + b \ln{r}}dx^{3}) \right)^{2} \right), 
& (D=4). \end{array} \right.  \label{tangmet} 
\end{eqnarray}
It is instructive to look at the two invariants $\tilde{R}$ and 
$\tilde{R}_{\mu\nu} \tilde{R}^{\mu\nu}$ at this point to see 
the singularity structure of the spacetimes described by
(\ref{tangmet}). In the simpler $D=4$ case, one finds that 
\[ \tilde{R} = \frac{b^{2} (1-2m) (1-2m+c_{1}^{2})}
{2r^{2}(a(1-2m) + b \ln{r})^{3}} \] 
and
\[ \tilde{R}_{\mu\nu} \, \tilde{R}^{\mu\nu} = 
\frac{b^{4} (1-2m)^{2} [3 (1-2m)^{2}+ 3 c_{1}^{4} -4(1-2m)c_{1}^{2}]}
{4r^{4}(a(1-2m) + b \ln{r})^{6}} . \]
Depending on how the integration constants $a$, $b$, $c_{1}$ and the mass
parameter $m$ are related to each other, one finds singularities at $r=0$ 
and at $r=r_{s}$ for which $\phi(r_{s}) \rightarrow \infty$, i.e. at
\[ r_{s} = e^{-a(1-2m)/b} > 0 \, . \] 
Similarly, even though there exists no explicit solution $u(r)$ of 
(\ref{ueqn}), one would expect to find singularities at $r=0$, at the
zeros of the metric function $\zeta(r)$ and at the $r$ values where 
$\phi(r) \rightarrow \infty$ when $D \ge 5$.  

It naturally follows that the conformally transformed G{\"o}del-type metric 
(\ref{tangmet}) can obviously be used as a solution to the supergravity 
theories that we have described in subsections \ref{6dim}, \ref{8dim} 
and \ref{10dim} by fixing the dimension $D$ accordingly. However one still 
needs to understand the physical relevance of the geometries described 
by (\ref{tangmet}) in the context of these theories.

\section{\label{ssd=2} $D=3$ solutions with two-dimensional backgrounds}

Let us now examine how things go for the special $D=3$ case separately.
When $D=3$, introduction of a 3-form field leads to a triviality since
any totally antisymmetric 3-tensor has to be proportional to the 
Levi-Civita tensor density, $\epsilon$, then. Hence we go back to
the Einstein tensor (\ref{emdeinst}) of the conformally transformed metric 
$\tilde{g}_{\mu\nu}$ and find that it simplifies to give
\begin{eqnarray}
\tilde{G}_{\mu\nu} & = & \frac{1}{2} \, \hat{r} \, \tilde{u}_{\mu} \, 
\tilde{u}_{\nu} + \frac{1}{2} \, \tilde{T}_{\mu\nu}^{\phi} 
+ \frac{1}{2} \, e^{-2 \phi} \, \tilde{T}_{\mu\nu}^{f} 
- \frac{1}{2} \, \tilde{g}_{\mu\nu} \left( \widetilde{\Box} \, \phi 
+ \frac{1}{2} \, e^{-2 \phi} \, \tilde{f}^2 
+ \tilde{u}_{\beta} \, \tilde{\nabla}_{\alpha} \,
(e^{-2 \phi} \, \tilde{f}^{\alpha\beta}) \right) \nonumber \\
& & + \frac{1}{2} \tilde{u}_{\mu} \left( \tilde{\nabla}_{\alpha} 
(e^{-2 \phi} \tilde{f}^{\alpha}\,_{\nu}) - \frac{1}{4}  
e^{-4 \phi} \tilde{f}^2 \tilde{u}_{\nu} \right)
+ \frac{1}{2} \tilde{u}_{\nu} \left( \tilde{\nabla}_{\alpha} 
(e^{-2 \phi} \tilde{f}^{\alpha}\,_{\mu}) - \frac{1}{4}  
e^{-4 \phi} \tilde{f}^2 \tilde{u}_{\mu} \right) \, ,
\label{d=3bas}
\end{eqnarray}
where we have used the fact that the Einstein tensor of a two-dimensional
metric vanishes identically.

At this stage there is no a priori way of choosing the Maxwell equation.
For simplicity though, let us assume that it is
\begin{equation}
\tilde{\nabla}_{\mu} (e^{-2 \phi} \, \tilde{f}^{\mu}\,_{\nu})
= a \, e^{-4 \phi} \, \tilde{f}^{2} \, \tilde{u}_{\nu} \, ,
\label{d=3max}
\end{equation}
where $a$ is an arbitrary function. In a manner similar to how we showed 
that (\ref{dilson}) follows from (\ref{maxson}), one can show 
that (\ref{d=3max}) implies
\begin{equation}
{\widetilde{\Box}} \, \phi = (a - \frac{1}{2}) \, e^{-2 \phi}  
\, \tilde{f}^{2} \, . \label{d=3dil}
\end{equation}
When (\ref{d=3max}) and (\ref{d=3dil}) are used in (\ref{d=3bas}), one gets
\begin{equation}
\tilde{G}_{\mu\nu} = \left( \frac{1}{2} \, \hat{r} + (a - \frac{1}{4}) 
\, e^{-4 \phi}  \, \tilde{f}^{2} \right) \tilde{u}_{\mu} \, \tilde{u}_{\nu}
+ \frac{1}{2} \, \tilde{T}_{\mu\nu}^{\phi} + \frac{1}{2} \, 
e^{-2 \phi} \, \tilde{T}_{\mu\nu}^{f} \, . \label{d=3ein}
\end{equation}
Together with (\ref{d=3max}) (and (\ref{d=3dil})), (\ref{d=3ein}) can
be interpreted as describing an Einstein-Maxwell-scalar field theory
coupled with a charged dust fluid in three dimensions with energy
density
\[ \rho = \frac{1}{2} \, \hat{r} + (a - \frac{1}{4}) 
\, e^{-4 \phi}  \, \tilde{f}^{2} \, . \]

Now let us take the special fixed coordinate $x^{k}$ as $x^{0} \equiv t$ 
without any loss of generality. It is well known that any two-dimensional 
metric is conformally flat and can be put in the form 
$h_{ij}=e^{2 \sigma(x,y)} \, \bar{\delta}_{ij}$. [Here $i,j=1,2$, we
take $x^{1} \equiv x$, $x^{2} \equiv y$ and in what follows we use
\[ \partial^{2} \sigma \equiv \bar{\delta}^{ij} \, \partial_{i} 
\, \partial_{j} \, \sigma \;\;\;\; \mbox{and} \;\;\;\; 
(\partial \sigma)^{2} \equiv \bar{\delta}^{ij} \, (\partial_{i} \, \sigma)
\, (\partial_{j} \, \sigma) \, . ] \]
Then the Ricci scalar of $h_{ij}$ is simply 
$\hat{r}=-2 \, e^{-2 \sigma} \, \partial^2 \sigma$. Let us furthermore
assume that
\[ u_{\mu} = e^{\phi(x,y)} \, \delta^{0}_{\mu} \, . \]

Put altogether, these assumptions lead to the conformally transformed
metric
\begin{equation}
ds^{2} = -dt^{2} + e^{2(\sigma-\phi)} \, (dx^{2} + dy^{2}) \, ,
\label{d=3met}
\end{equation}
and \( f_{\mu\nu} = (\delta_{\mu}^{i} \, \delta_{\nu}^{0} - 
\delta_{\mu}^{0} \, \delta_{\nu}^{i}) \, (\partial_{i} \phi)
\, e^{\phi} \, . \)
Now a careful calculation gives 
\( \tilde{f}^{2}=-2 \, (\partial \phi)^{2} \, e^{4\phi - 2\sigma} \, . \) 
This in turn implies that 
\( \rho = \left( (\frac{1}{2} -2a) \, (\partial \phi)^{2} 
- \partial^{2} \sigma \right) \, e^{-2 \sigma} \, . \)
Using these, one finds that (\ref{d=3max}) (or equivalently (\ref{d=3dil}))
is satisfied provided
\begin{equation} 
\partial^{2} \phi = (1-2a) \, (\partial \phi)^{2} \, . \label{integ}
\end{equation}
For constant $a$, this means that \( \partial^{2} (e^{(2a-1)\phi}) =0 \) 
when $a \neq 1/2$ and $\partial^{2} \phi=0$ when $a=1/2$. Therefore any 
function $\phi(x,y)$ for which $e^{(2a-1)\phi}$ is harmonic in the $(x,y)$ 
variables solves (\ref{integ}) when $a$ is a constant and $a \neq 1/2$. 
When $a=1/2$, $\phi$ itself must be a harmonic function.

So given the value of $a$ and $\rho$, one can first find a suitable
$\phi$ and using this, determine the function $\sigma(x,y)$ to construct 
the metric (\ref{d=3met}) that exactly solves the theory described by
equations (\ref{d=3ein}), (\ref{d=3max}) and (\ref{d=3dil}).

As a specific example, suppose that there is no dust fluid present, i.e.
the energy density $\rho$ has been set to zero. Now using (\ref{integ})
in the expression for $\rho$, one finds that
\begin{equation} 
\partial^{2} \sigma = \frac{1-4a}{2(1-2a)} \, \partial^{2} \phi \qquad 
\mbox{for} \quad a \neq \frac{1}{2} \, , \label{sip}
\end{equation}
and given $a \neq 1/2$, one can first determine $\phi$ through 
(\ref{integ}) and then use this in (\ref{sip}) to determine $\sigma$.

\section{\label{conc} Conclusions}

We have used the previously introduced G{\"o}del-type metrics to find 
solutions to the Einstein field equations coupled with a Maxwell-dilaton-3-form
field theory in $D \geq 6$ dimensions. By construction the matter fields
were related to the vector field $u_{\mu}$ and their field equations were
shown to follow from the ``Maxwell equation'' for $u_{\mu}$. We showed that
there is a vast number of possibilities for the choice of the 
$(D-1)$-dimensional Riemannian spacetime backgrounds $h_{\mu\nu}$ and that
one can find exact solutions to the bosonic field equations of supergravity 
theories in six, eight and ten dimensions by effectively reducing these
equations to a single ``Maxwell equation'' (\ref{maxinh}) in the relevant
background $h_{\mu\nu}$. Specifically when $h_{\mu\nu}$ is the 
$(D-1)$-dimensional flat Riemannian geometry, the solutions found happen
to be the $D$-dimensional generalizations of the $D=4$ dimensional 
Majumdar-Papapetrou metrics. The $D=3$ case was also shown to admit
a family of solutions describing a Maxwell-dilaton field.

It would be worth studying to seek other theories for which the techniques
we have employed here are applicable. A detailed analysis of the solutions
we have found and an investigation of how much supersymmetry, if any, they
preserve is one possible future direction to look at. It would also be very
interesting to try out more general $(D-1)$-dimensional Riemannian 
spacetimes (some of which we have articulated in the second to last
paragraph of section \ref{emdil}) as backgrounds, to solve (\ref{maxinh}) 
using these and to find possibly new solutions. Another attractive avenue
to look at would be to try out G{\"o}del-type metrics for which
$\partial_{k} g_{\mu\nu} \neq 0$, contrary to what has been assumed so far.

\section{\label{acknow} Acknowledgments}

We would like to thank Atalay Karasu for his participation in the early
stages of this work and for a careful reading of this manuscript. We would
also like to thank the referees for their diligence. This work is partially 
supported by the Scientific and Technical Research Council of Turkey and 
by the Turkish Academy of Sciences.

\appendix
\section{\label{appa} Preliminaries for obtaining (\ref{emdein})}

In this appendix we present some calculations which are useful 
in the derivation of (\ref{emdein}).

By definition,
\[ \nabla_{\beta} \, u_{\alpha} = \partial_{\beta} u_{\alpha} 
- \Gamma^{\mu}\,_{\alpha\beta} \, u_{\mu} 
= u_{\alpha \vert \beta} 
- C^{\mu}\,_{\alpha\beta} \, u_{\mu} \, , \]
where we have introduced 
\[ C^{\mu}\,_{\alpha\beta} \equiv 
\Gamma^{\mu}\,_{\alpha\beta} - \gamma^{\mu}\,_{\alpha\beta} = \frac{1}{2}\, 
(u_{\alpha} \, f^{\mu}\,_{\beta} + u_{\beta} \, f^{\mu}\,_{\alpha}) 
- \frac{1}{2} \, u^{\mu} \, 
(u_{\alpha \vert \beta} + u_{\beta \vert \alpha}) \, . \]

By defining $\phi \equiv \ln{|u_{k}|}$ and denoting 
$\phi_{\mu} = \nabla_{\mu} \, \phi$,
one can show that the following identities hold:
\begin{eqnarray} 
u^{\alpha} \, \nabla_{\beta} \, u_{\alpha} & = & 0 \, , \label{bir} \\
u^{\mu} \, f_{\mu \nu} & = & \phi_{\nu} \, , \label{iki} \\
u^{\mu} \, \phi_{\mu} & = & 0 \, . \label{uc} 
\end{eqnarray}
The first one is found by using $u_{\mu} u^{\mu}= -1$ 
whereas the next two follow from the form of $u^{\mu}$, the assumption
that $u_{\mu}$ is independent of $x^{k}$ and the definition of $\phi$.
Similarly one obtains the following formulas regarding the 
derivatives of $u_{\mu}$:
\begin{eqnarray}
\nabla_{\alpha} \, u_{\beta} & = & \frac{1}{2} \,[ f_{\alpha\beta}
- u_{\alpha} \, \phi_{\beta} - u_{\beta} \, \phi_{\alpha} ] \, , 
\label{dort} \\
\nabla_{\alpha} \, u^{\alpha} & = & 0 \, , \label{bes} \\
u^{\alpha} \, \nabla_{\alpha} \, u_{\beta} & = & \phi_{\beta} \, , 
\label{alti} \\
\partial_{\alpha} \, u^{\alpha} & = & 0 \, . \label{yedi}
\end{eqnarray} 
(\ref{dort}) is found by explicitly writing the left hand side in terms of 
the Christoffel symbols and using (\ref{iki}). (\ref{bes}) and (\ref{alti}) 
follow from (\ref{dort}), (\ref{iki}) and (\ref{uc}). Finally (\ref{yedi}) 
is obtained by using $u^{\mu}$ explicitly and the assumption that all 
quantities are independent of $x^{k}$. It immediately follows that, unlike 
the case for constant $u_{k}$ examined in \cite{gks}, $u^{\mu}$ is no longer 
a timelike Killing vector due to (\ref{dort}) and is no longer tangent to 
a timelike geodesic curve by (\ref{alti}).

Moreover, the following identities regarding the Christoffel
symbols are useful in the calculation of the Ricci tensor: 
\begin{eqnarray}
\gamma^{\nu}\,_{k\alpha} & = & 0 \, , \label{sekiz} \\
u^{\mu} \, \gamma^{\nu}\,_{\mu\alpha} & = & 0 \, , \label{dokuz} \\
C^{\nu}\,_{\nu \mu} & = & \phi_{\mu} \, , \label{on} \\
\Gamma^{\nu}\,_{\nu \mu} & = & \phi_{\mu} +
\gamma^{\nu}\,_{\nu\mu} \, . \label{onbir} 
\end{eqnarray}
(\ref{sekiz}) and (\ref{dokuz}) again result from the independence 
of all quantities from $x^{k}$ and the forms of $u^{\mu}$ and $h_{k\mu}$. 
Since \( \sqrt{-g}=|u_{k}| \, \sqrt{|h|} \), one finally gets (\ref{on}) 
and (\ref{onbir}) by the definition of $\phi$. 

With the help of these identities, the Ricci tensor takes the form:
\begin{equation}
R_{\mu \nu} = \hat{r}_{\mu\nu} + \frac{1}{2} \, f_{\mu}\,^{\alpha} 
\, f_{\nu \alpha} + \frac{1}{4} \, f^2 \, u_{\mu} \, u_{\nu}
- \nabla_{\mu} \nabla_{\nu} \phi - \frac{1}{2} \, \phi_{\mu} \, \phi_{\nu}  
+\frac{1}{2} \, (u_{\mu} \, j_{\nu} + u_{\nu} \, j_{\mu}) \,
\label{emdric1}
\end{equation}
where we use 
\[ f^2 \equiv f^{\alpha \beta} \, f_{\alpha \beta} \, , \;\;\;\;
j_{\mu} \equiv f^{\alpha}\,_{\mu \vert \alpha} 
- \phi^{\alpha} \, u_{\mu \vert \alpha} \] 
and $\hat{r}_{\mu\nu}$ denotes the Ricci tensor of 
$\gamma^{\mu}\,_{\alpha\beta}$. However for later convenience 
it is better to convert the covariant derivatives with respect to 
$\gamma^{\mu}\,_{\alpha\beta}$ in $j_{\mu}$ to covariant derivatives with 
respect to $\Gamma^{\mu}\,_{\alpha\beta}$. This is achieved by noting that
\[ j_{\mu} = \nabla_{\alpha} \, f^{\alpha}\,_{\mu} - \phi_{\alpha} \, 
f^{\alpha}\,_{\mu} - \frac{1}{2} \, f^2 \, u_{\mu} = e^{\phi} \, 
\nabla_{\alpha} \, (e^{-\phi} \, f^{\alpha}\,_{\mu}) - \frac{1}{2} 
\, f^2 \, u_{\mu} \]
and using this in (\ref{emdric1}), the Ricci tensor now becomes
\begin{eqnarray}
R_{\mu \nu} & = & \hat{r}_{\mu\nu} 
+ \frac{1}{2} \, u_{\mu} \, e^{\phi} \, \nabla_{\alpha} \, 
(e^{-\phi} \, f^{\alpha}\,_{\nu}) + \frac{1}{2} 
\, u_{\nu} \, e^{\phi} \, \nabla_{\alpha} 
\, (e^{-\phi} \, f^{\alpha}\,_{\mu}) \nonumber \\ 
& & + \frac{1}{2} \, f_{\mu}\,^{\alpha} \, f_{\nu \alpha} 
- \frac{1}{4} \, f^2 \, u_{\mu} \, u_{\nu}
- \nabla_{\mu} \nabla_{\nu} \phi 
- \frac{1}{2} \, \phi_{\mu} \, \phi_{\nu} \, .
\label{emdricci}
\end{eqnarray}
The Ricci scalar is then given by 
\begin{equation}
R = \hat{r} + u_{\nu} \, e^{\phi} \, \nabla_{\mu} \, (e^{-\phi} \, f^{\mu\nu})
+ \frac{3}{4} \, f^2 - \Box \, \phi - \frac{1}{2} \, (\nabla \, \phi)^{2} \, ,
\label{emdricsca}
\end{equation}
where we use 
\[ \Box \, \phi \equiv g^{\mu\nu} \nabla_{\mu} \nabla_{\nu} 
\phi \, , \;\;\;\;
(\nabla \, \phi)^{2} \equiv g^{\mu\nu} (\nabla_{\mu} \phi) 
(\nabla_{\nu} \phi) \, , \;\;\;\;
f^{\mu\nu} \equiv g^{\mu\alpha} g^{\nu\beta} f_{\alpha\beta} \, , \]
and $\hat{r}$ denotes the Ricci scalar of $\gamma^{\mu}\,_{\alpha\beta}$.
Note that 
\[ \hat{r} = g^{\alpha\beta} \, \hat{r}_{\alpha\beta} = 
\bar{h}^{\alpha\beta} \, \hat{r}_{\alpha\beta} \]
by using $u^{\mu}=-\frac{1}{u_{k}} \delta^{\mu}_{k}$, 
(\ref{invmet}) and (\ref{dokuz}) in the explicit calculation of $\hat{r}$.
Now using (\ref{emdricci}) and (\ref{emdricsca}), one obtains (\ref{emdein})
in the text.

\section{\label{appb} Getting (\ref{emdeinst}) from (\ref{emdeintil})}

In this appendix we present the details of the calculations 
that lead to (\ref{emdeinst}).

The first task is to relate the covariant derivatives $\tilde{\nabla}_{\mu}$ 
and $\nabla_{\mu}$ to each other. This is given by \cite{wald}
\[ \tilde{\nabla}_{\mu} \, \omega_{\nu} = \nabla_{\mu} \, \omega_{\nu} 
- K^{\sigma}\,_{\mu\nu} \, \omega_{\sigma} \, , \]
where $\omega_{\mu}$ is any vector field and
\[ K^{\sigma}\,_{\mu\nu} = \frac{1}{2-D} \left( \delta^{\sigma}_{\;\mu} \, 
\nabla_{\nu} \phi + \delta^{\sigma}_{\;\nu} \, \nabla_{\mu} \phi 
-g_{\mu\nu} \, g^{\sigma\alpha} \, \nabla_{\alpha} \phi \right) \]
in our case. We now assume that the vector $u_{\mu}$ is conformally 
invariant with conformal weight zero, i.e. $\tilde{u}_{\mu}=u_{\mu}$. 
Obviously, one also has 
$\tilde{\nabla}_{\mu} \, \phi \equiv \nabla_{\mu} \phi$ 
then. Using these and defining \( \tilde{f}_{\alpha\beta} = 
\tilde{\nabla}_{\alpha} \, \tilde{u}_{\beta} - \tilde{\nabla}_{\beta} \, 
\tilde{u}_{\alpha} \, ,\) it readily follows that the ``Maxwell field'' 
$f_{\alpha\beta}$ also has conformal weight equal to zero, 
i.e. $\tilde{f}_{\alpha\beta}=f_{\alpha\beta}$. 

At this point special care has to be given as to which metric is used for
raising and lowering indices. To remove any ambiguity, we now assume that
any quantity with a tilde on it has all its indices raised and lowered by
using the metric $\tilde{g}_{\mu\nu}$ only; thus for example one has 
\[ \tilde{f}^{\mu}\,_{\nu} = \tilde{g}^{\mu\alpha} \, \tilde{f}_{\alpha\nu} 
= e^{\frac{2 \phi}{D-2}} \, g^{\mu\alpha} \, f_{\alpha\nu} = 
e^{\frac{2 \phi}{D-2}} \, f^{\mu}\,_{\nu} \]
in our case. One then finds the following useful identities all of which
follow from the preliminaries above:
\begin{eqnarray*}
f^2 & = & e^{\frac{4 \phi}{2-D}} \, \tilde{f}^2 \, , \\
T_{\mu\nu}^{f} & = & e^{\frac{2 \phi}{2-D}} \, \tilde{T}_{\mu\nu}^{f} \, , \\
T_{\mu\nu}^{\phi} & = & \tilde{T}_{\mu\nu}^{\phi} \, , \\
\nabla_{\mu} \nabla_{\nu} \phi & = & \tilde{\nabla}_{\mu} \, 
\tilde{\nabla}_{\nu} \, \phi 
+ \frac{2}{2-D} \, \tilde{T}_{\mu\nu}^{\phi} \, , \\
g_{\mu\nu} \, \Box \, \phi & = & \tilde{g}_{\mu\nu} \, 
({\widetilde{\Box}} \, \phi + ( \tilde{\nabla} \, \phi )^2) \, , \\
g_{\mu\nu} \, g^{\alpha\beta} & = & \tilde{g}_{\mu\nu} \, 
\tilde{g}^{\alpha\beta} \, , \\
e^{\phi} \, \nabla_{\mu} \, (e^{-\phi} \, f^{\mu}\,_{\nu}) & = & 
\tilde{\nabla}_{\mu} \,
(e^{\frac{2 \phi}{2-D}} \, \tilde{f}^{\mu}\,_{\nu}) \, , \\
u_{\nu} \, e^{\phi} \, \nabla_{\mu} \, (e^{-\phi} \, f^{\mu\nu}) & = & 
\tilde{u}_{\nu} \, e^{\frac{2 \phi}{2-D}} \, \tilde{\nabla}_{\mu} \,
(e^{\frac{2 \phi}{2-D}} \, \tilde{f}^{\mu\nu}) \, .
\end{eqnarray*}

Using these in (\ref{emdeintil}) and arranging the resulting 
terms carefully, one obtains the Einstein tensor given in (\ref{emdeinst}) 
which now only involves terms associated with the conformally transformed 
metric (or the string metric) $\tilde{g}_{\mu\nu}$.

\end{document}